\documentclass[aip, prl, reprint]{revtex4-1}

\usepackage[breaklinks, hidelinks]{hyperref}
\usepackage{graphicx}
\usepackage{caption}
\usepackage{subcaption}
\usepackage{amsmath}
\usepackage{amsfonts}
\usepackage{fancyhdr}
\usepackage{ulem}
\usepackage{cancel}
\usepackage{natbib}
\usepackage{color}

\usepackage[table]{xcolor}

\begin{document}
	\title{Citation histories of papers: sometimes the rich get richer, sometimes they don't}
	
	\author{Michael J. Hazoglou}
	\affiliation{Department of Physics \& Astronomy, Stony Brook University}
	\author{Vivek Kulkarni}
	\affiliation{Department of Computer Science, Stony Brook University}
	\author{Steven S. Skiena}
	\affiliation{Department of Computer Science, Stony Brook University}
	\author{Ken A. Dill}
	\affiliation{Department of Physics \& Astronomy, Stony Brook University}
	\affiliation{Laufer Center for Physical and Quantitative Biology}
	\affiliation{Department of Chemistry, Stony Brook University}

	\begin{abstract}
		We describe a simple model of how a publication's citations change over time, based on pure-birth stochastic processes with a linear cumulative advantage effect. The model is applied to citation data from the Physical Review corpus provided by APS. Our model reveals that papers fall into three different clusters: papers that have rapid initial citations and ultimately high impact (\textit{fast-hi}), fast to rise but quick to plateau (\textit{fast-flat}), or late bloomers (\textit{slow-late}), which may either never achieve many citations, or do so many years after publication.  In \textit{fast-hi} and \textit{slow-late}, there is a rich-get-richer effect: papers that have many citations accumulate additional citations more rapidly while the \textit{fast-flat} papers do not display this effect.  We conclude by showing that only a few years of post-publication statistics are needed to identify high impact (\textit{fast-hi}) papers.
	\end{abstract}
	
	\maketitle

	\section{Introduction}
	Every research publication has a citation history.  A paper's citation history contains more information about the paper than any single-number measure alone does, such as its citation count. Here, we develop a generative model to describe the citation histories of publications, and we ask how predictable a paper's future citations are from its initial citations (citation history).
	
	We begin with a few general observations about citations.  Like a person, a publication goes through birth-and-death stages.  Soon after it is `born', the paper receives some initial number of citations per year (its citation rate).  Later, when interest in the paper has waned, the citation rate diminishes toward zero.  Certain papers have high citation rates; others have low rates.  Some papers continue receiving citations over a long lifetime, while others die out quickly.  Interestingly, one of the earliest papers in the field of bibliometrics, by Lotka in 1926, leading to what is now called `Lotka's Law \cite{Lotka1926,ChungAndCox1990}, is itself still quite highly cited.

	When looking at multiple papers collected together as a set -- for example, the collected papers of an individual or over a field of research -- it is found that a plot of the numbers of papers as a function of the citation counts of those papers shows distributions shaped like a power-law \cite{Newman2006} or like the exponential of a digamma function \cite{Peterson2010}.  The non-exponential tails of these distributions indicate a `rich-get-richer' feature: the more citations a paper already has, the higher the rate at which it receives new citations.  
	
	These features are captured in a recent model (WSB) of the time-dependent histories of citations over time~\cite{WangSongBarabasi2013,Shen2014,Xiao2016}.  Here, we go beyond that model in two ways.  First, we provide a more microscopic, or generative, model for the underpinnings of citations.  Second, we introduce a different but related mechanistic model that we call the \textit{Direct-Indirect} (DI) model \cite{Peterson2010}, which has a simple mechanistic interpretation and captures observed data with fewer unexplained parameters.  We use our models to analyze a set of $151,082$ papers from the Physical Review Corpus of the American Physical Society across all the disciplines represented by papers published by APS journals. The model is then used to classify papers into particular categories of performance based on the similarity of the papers.

	\section{The Pure Birth Process and the Mathematics of a Process with Cumulative Advantage}
	In this section we review a special case of birth-death processes, the pure birth process. Birth-death processes involve an increase in the state variable by one, called birth and a decrease in the state variable by one called death. Here the state variable will be the total number of citations a paper has accumulated. The pure birth process is useful in bibliometrics citations because papers generally don't disappear (retractions are negligibly rare events).  A pure birth process resembles a process of compounded interest; the formalism below follows from that.  This formalism allows the user a lot of flexibility in describing bibliometric data: as long as the four defining properties of the birth process are true the rate can be refined with different functional forms to model the data. The birth process is described in terms of a conditional probability distribution $p(N(t+\tau)-N(t) = n| N(t)=m) \equiv p_{m,n}(t,\tau)$ for the number of events $N(t)$ at the time $t$.  The initial condition is that $N(0) = 0$.  We assume the following properties \cite{panjer1992insurance}:
	
	\begin{itemize}
		\item The process is Markovian depending only on the current state $N(t), t \geq 0$
		\item $p_{m,1}(t,\delta t)= \lambda_{m}(t) \delta t + o(\delta t)$ for $m=0, 1, 2, 3, \dots$
		\item $p_{m,n}(t,\delta t)=  o(\delta t)$ for $n>1$, $m=0, 1, 2, 3, \dots$
	\end{itemize}
	here $\lambda_{m}(t)$ is the rate which can depend on the total number of events and the time. These conditions can be used to define a set of differential equations to solve for $p_{m,n}(t,\tau)$ \cite{panjer1992insurance}, but a general solution can be obtained in terms of a recursive integral equation, which is simpler and straight forward to understand. Since cumulative advantage is important in citation dynamics we will focus on processes with rates $\lambda_n (t) = (a n + b)\lambda(t)$ with the probability distribution
	\begin{equation}
		p_{m,n}(t,\tau) = \frac{\Gamma(b/a + m +n)}{n! \Gamma(b/a + m)}[e^{-a\gamma(t,\tau)}]^{\frac{b}{a}+m}[1-e^{-a\gamma(t,\tau)}]^n
	\label{NegativeBinomialDistributionBirthProcess}
	\end{equation}
	where $\gamma(t,\tau) = \int_{t}^{\tau+t} \lambda(u) du$, and $\Gamma(x+1) = x \Gamma(x)$ is the standard gamma function.
	
	The mean number of events in the time interval $\tau$ given $m$ events in the time $t$ is given by
	\begin{equation}
		\left\langle n(\tau)|m,t\right\rangle = \left(\frac{b}{a}+m\right)\left[e^{a\gamma(t,\tau)}-1\right].
		\label{MeanEventsBirthProcess}
	\end{equation}
		
	Eqn.~\eqref{NegativeBinomialDistributionBirthProcess} is negative binomial distribution, which is characteristic of a process with cumulative advantage where the frequency of events increases with the number of events. In the context of citations, this would model the discovery of a paper through the references of another more recent paper. The limit as $a \rightarrow 0^+$ describes the situation in which there is no cumulative advantage, and it reduces to a Poisson distribution.
	
	It is worth noting that the main results of Shen, Wang, Song, and Barab\'{a}si \cite{WangSongBarabasi2013, Shen2014} are derived from substituting into Eqn.~\eqref{MeanEventsBirthProcess} with the choice of $\lambda(t)$ being the same functional form as a log-normal distribution. What was not obtained in those papers was a probability distribution for the process. As a result the process has been mislabeled as a ``reinforced Poisson process'' when in fact Eqn.~\eqref{NegativeBinomialDistributionBirthProcess} is a negative binomial distribution \cite{Shen2014}.

	\section{Direct and Indirect Mechanism of Citation}
	\label{ExpModelSec}
	Here, we consider the Direct-Indirect (DI) mechanism of citations \cite{Peterson2010}.  In the DI mechanism, a paper $B$ receives a \textit{direct citation} from a paper $A$ when the author of $A$ finds $B$ through some direct search over a database of papers.  In contrast, a paper $B$ receives an \textit{indirect citation} when the author of $A$ finds paper $B$ in the references of another paper. The difference in these mechanisms is that of being searched and found.  A direct search finds the single paper $B$.  An indirect search finds any one of many different papers that cite $B$.  We model the rate $\lambda_n(t)$ at which a paper having $n$ citations is cited at time $t$ as:
		
	\begin{equation}
	\lambda_n(t)= \lambda_{\mbox{indirect}}(t)+\lambda_{\mbox{direct}}(t)
	\end{equation}
	If we say a paper is found selected almost randomly from a body of relevant papers, then
	\begin{equation}
	\lambda_{\mbox{direct}}(t) \propto \frac{1}{N(t)}
	\end{equation}
	$N(t)$ is the number of relevant papers at time $t$, and the indirect rate is proportional to the number of citations a paper has received at a given time $n(t)$, so
	
	\begin{equation}
	\lambda_{\mbox{indirect}}(t) \propto \frac{n(t)}{N(t)}
	\end{equation}
since the number of papers increases exponentially with time $N(t) \propto \exp(r t)$ \cite{Larsen2010, WangSongBarabasi2013}.
Therefore, we obtain:
	\begin{equation}
	\lambda_n(t) =(a n + b) r \exp(-r t).
	\end{equation}
	
	This choice of rate gives $\gamma(t,\tau) = \exp(-r t)- \exp(-r(t+\tau))$, the mean number of citations from Eq.~\eqref{MeanEventsBirthProcess}, with $t=0$ and $m=0$ as
	
	\begin{equation}
	\left\langle n(\tau)|0,0\right\rangle = \frac{b}{a}\left[\exp\left(a(1-e^{-r \tau})\right)-1\right]
	\label{ExpModelMean}
	\end{equation}
	$b r$ is easily interpreted as the initial citation and $a r$ is how much the rate increases with each citation the paper receives within a short time ($\tau \ll 1/r$).
	
	Over long time ranges, we can use this to determine the expected number of new citations a paper will receive given some $m$ and $t$:
	\begin{equation}
	\lim_{\tau \rightarrow \infty} \left\langle n(\tau)|m,t\right\rangle = \left(\frac{b}{a}+m\right)\left[\exp\left(a e^{-r t}\right)-1\right]
	\end{equation}
	Setting $m=0$ and $t=0$ gives the expected number of total citations
	\begin{equation}
	\lim_{\tau \rightarrow \infty} \left\langle n(\tau)|0,0\right\rangle = \frac{b}{a}\left[\exp\left(a\right)-1\right]
	\end{equation}
	which for the limiting case of $a=0$ reduces to $b$. This expression tells us according to this model that the expected total citations grows exponentially with the cumulative advantage parameter $a$. Even moderately large values of $a$ would result in unreasonably large values unless the parameter $b$ exponentially decreases with $a$.
	
	The relevant time scale for this process is $1/r$.
With our original motivation, this will give us the doubling time from
	\begin{equation}
	t_{2\times} = \frac{\ln(2)}{r}
	\end{equation}
	where a typical doubling time of 13 years would result in a $r\approx 0.05 \, \mbox{year}^{-1}$.
	
	A Taylor series expansion around $\tau=0$ of $\langle n(\tau) | 0, 0\rangle$ gives the expected expected initial rate of citation, which will be called the \textit{velocity} $v(0)$.
	
	\begin{equation}
		v(0) \equiv \left. \frac{d}{d\tau} \langle n(\tau)|0, 0\rangle \right|_{\tau=0} = br
	\end{equation}

	\section{Clustering the model parameters shows that most papers fall into one of three classes}
	
	We apply this model to a database of 151,082 papers from the Physical Review journals.  For each paper, we extract the three parameters of the model: $b$ (proportional to the initial citation velocity $br$), $a$ (a `rich-get-richer' parameter), and $r$, a characteristic inverse time scale.  We then cluster papers together in terms of those three parameters using the clustering method called DBSCAN. 
	
\begin{figure}
		\begin{subfigure}{\linewidth}
			\includegraphics[width=\linewidth]{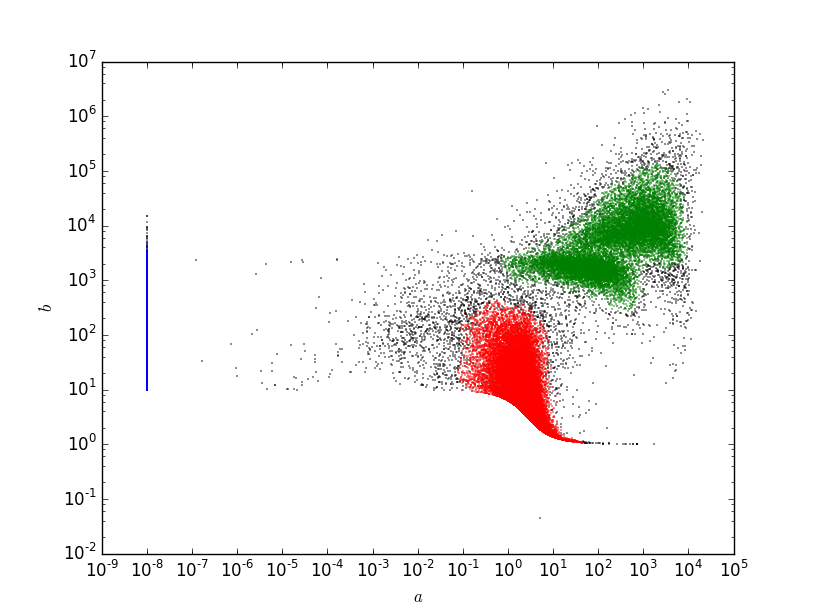}
			\caption{$b$ vs. $a$}
			\label{ExpClusters_b_vs_a}
		\end{subfigure}
		
		\begin{subfigure}{\linewidth}
			\includegraphics[width=\linewidth]{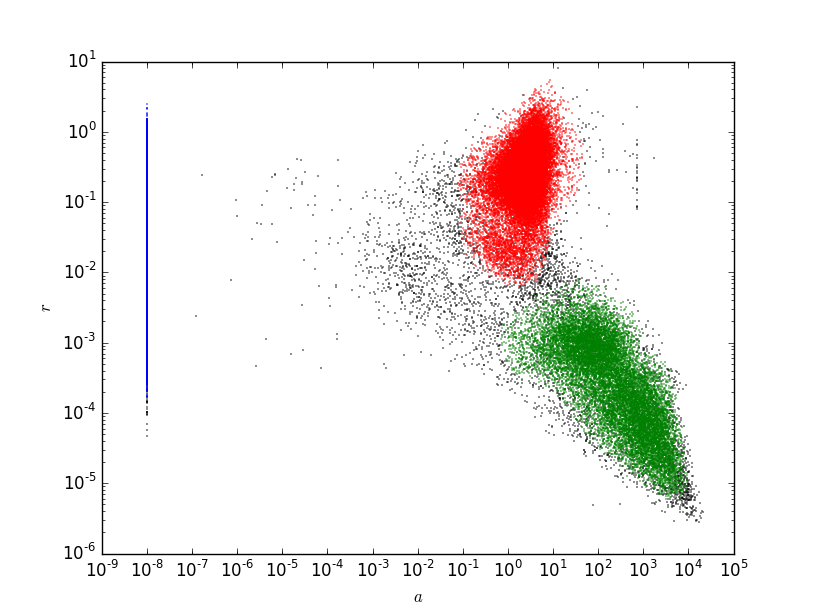}
			\caption{$r$ vs. $a$}
			\label{ExpClusters_r_vs_a}
		\end{subfigure}
		
		\begin{subfigure}{\linewidth}
			\includegraphics[width=\linewidth]{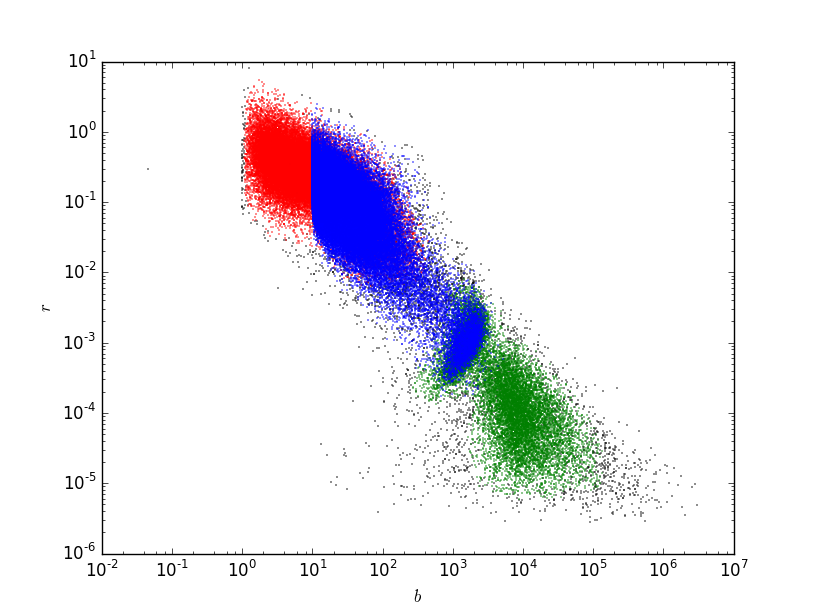}
			\caption{$r$ vs. $b$}
			\label{ExpClusters_r_vs_b}
		\end{subfigure}
		\caption[Scatter plot showing the clustering of papers based on DBSCAN projected onto planes.]{Scatter plot showing the clustering of papers based on the DBSCAN algorithm,  with $\epsilon = 0.5$ and a minimum threshold of 50 papers projected onto (a) the $b$ and $a$ axes, (b) the $r$ and $a$ axes, and (c) the $r$ and $b$ axes. The different clusters are color coded as red (\textit{fast-hi}), green (\textit{slow-late}), and blue (\textit{fast-flat}). The black points are papers categorized as noise. The parameter $a$ is bounded from below by $10^{-8}$.}
		\label{ExpClusters}
	\end{figure}

\begin{figure}
		\centering
		\includegraphics[width=\linewidth]{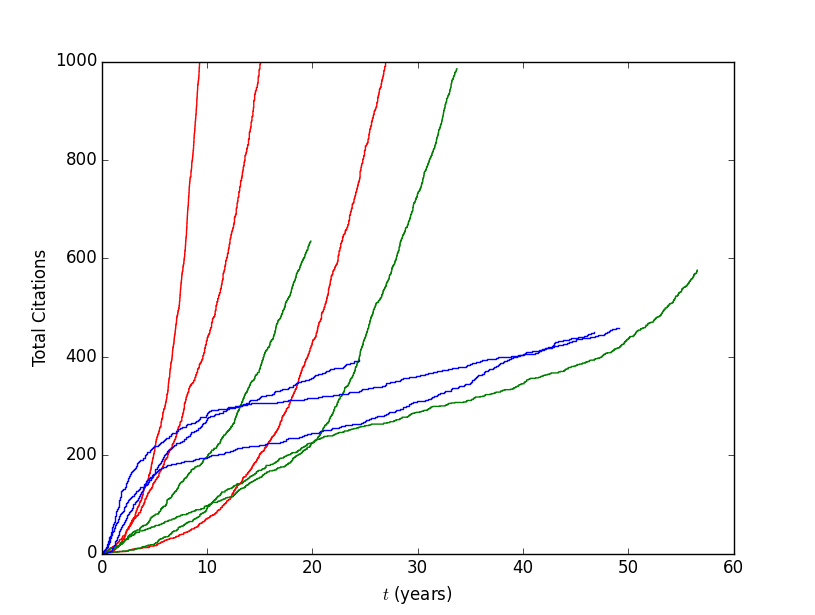}
		\caption[Citations trajectories of the ten most cited papers in each cluster]{Citation trajectories of the ten most cited papers in each cluster. The colors indicate the clusters the papers belong to and are the same as Figure~\ref{ExpClusters}}
		\label{CitationTrendsOfTop10perCluster}
	\end{figure}

	 Figure~\ref{ExpClusters} shows two plots of different parameter pairs, showing that the model clusters papers into three main categories:  the \textit{fast and high-impact} \textit{fast-hi} (shown in red) has a good initial velocity, benefits from a favorable rich-get-richer contribution, and has a long decay time, the \textit{quick to plateau} \textit{fast-flat} papers (blue) do not benefit from a rich-get-richer component.  Finally, \textit{slow-late} papers (green) tend to be long-lived and to benefit from acceleration after a long delay; they are sleepers. 
	 Examples of the three corresponding time trajectories are shown in Figure~\ref{CitationTrendsOfTop10perCluster}.

\begin{figure}
		\begin{subfigure}{.5\textwidth}
			\includegraphics[width=\linewidth]{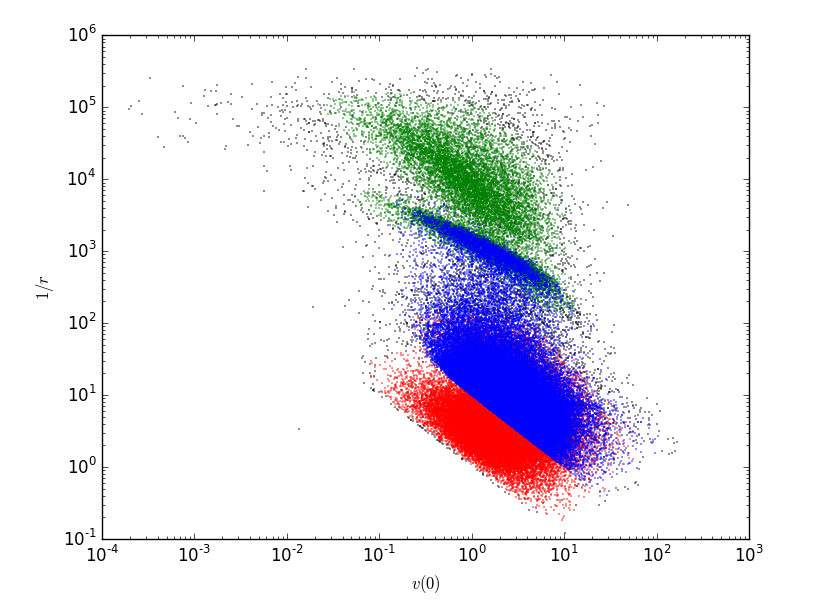}
			\caption{}
			\label{1overrVsVelocity}
		\end{subfigure}
		\begin{subfigure}{.5\textwidth}
			\includegraphics[width=\linewidth]{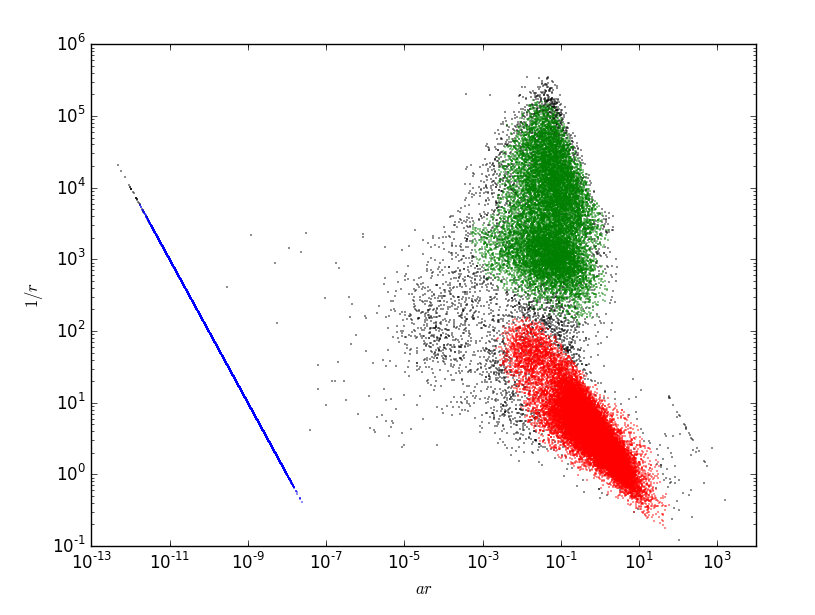}
			\caption{}
			\label{1overrVsar}
		\end{subfigure}
		\caption{We compare (a) the characteristic time scale $1/r$ versus expected initial velocity, and (b) the characteristic time scale $1/r$ versus the short term increase in citation rate per citation $a r$. The colors indicate the same clusters
as Figure~\ref{ExpClusters}.}
		\label{SeparationClusters}
	\end{figure}

	 Figure~\ref{SeparationClusters} makes two points. First, it shows that the initial citation rate $v(0)$ and the decay rate $r$ of a paper are independent parameters. One cannot be predicted from knowing the other. Second, it shows that the difference in $a r$ between the \textit{fast-hi} and the \textit{slow-late} clusters are independent of $r$. The quantity $a r$ is a measure of how the rate increases upon citation for times scales less than $1/r$.
	 Since $1/r$ is a characteristic time scale for the citation, we  see in Figure~\ref{1overrVsVelocity} that the initial expected rate of citation is not a strong indicator of the longevity of a paper. Figure~\ref{1overrVsar} shows the same separation from Figure~\ref{ExpClusters} in terms of physical quantities, as $ar$ is the initial increase in citation rate every time the paper is cited.
	 
	\begin{figure}
		\centering
		\includegraphics[width=\linewidth]{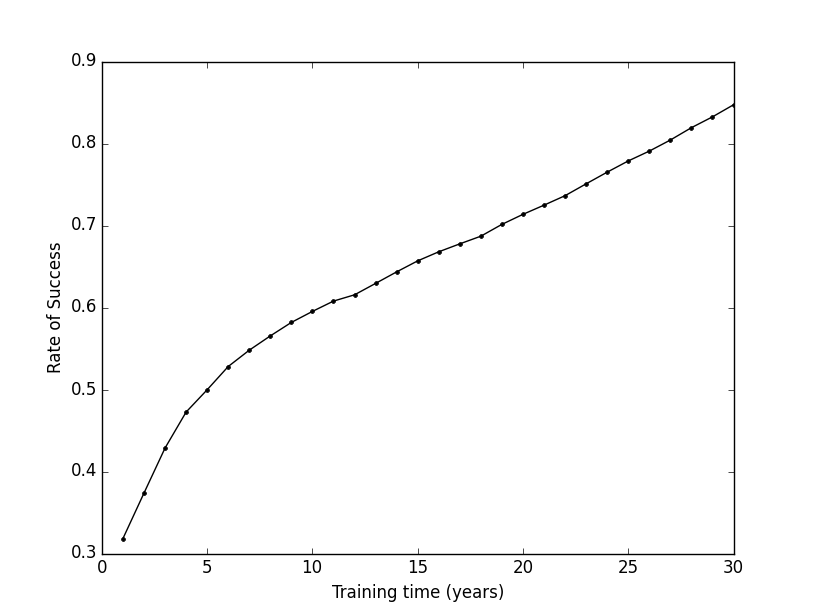}
		\caption[Success Rate and Median Age vs Training Time]{Success Rate versus training time in years for the all classifiable papers over the age of 30 years.}
		\label{SuccessRateVsTraining}
	\end{figure}
	
	\section{Can we predict a full citation history of a paper from its first few years?}
	
	How well can we estimate the three parameters of a paper's citation history, after observing only the first few years following publication?  First, we consider a three-category classification problem. At a given year after publication, which we call the observation time, what is the fraction of papers more than 30 years old that are correctly classified into one of the three clusters using only the data from this training period? We note that for a 3-category classification, purely random chance will give us a 33\% success rate. Figure~\ref{SuccessRateVsTraining} shows the prediction success as a function of training time. Success rises from 33\%, reaches 50\% at 5 years of training time, and continues increasing monotonically with increasing observation time.
	
\begin{figure}
		\centering
		\includegraphics[width=\linewidth]{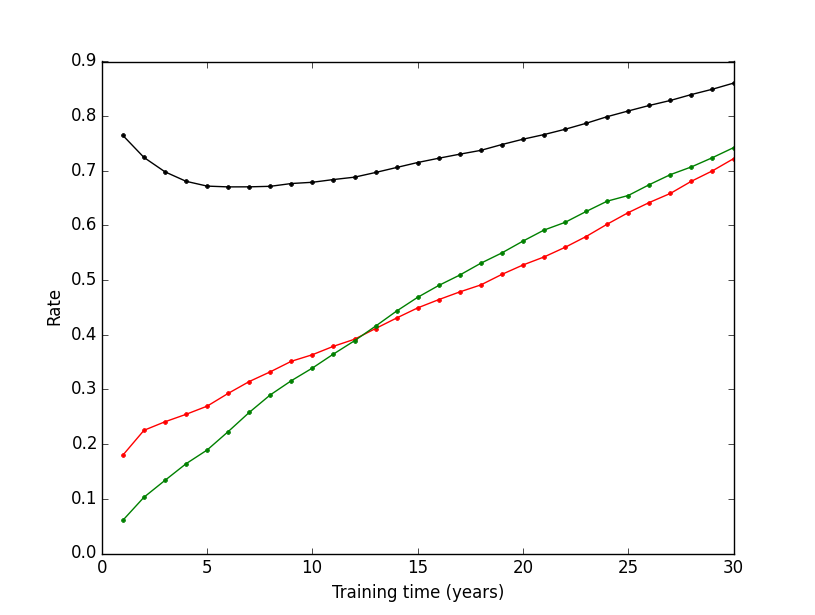}
		\caption[Binary Success Rate and Median Age vs Training Time]{Success Rate versus training time in years for the binary classification of \textit{fast-hi} or not \textit{fast-hi} in black, the precision (positive predictive value) in red and the recall (true positive rate) in green for all papers over 30 years of age.}
		\label{SuccessRateVsTrainingBinary}
	\end{figure}

	Second, we also consider a 2-category classification problem. Here we seek to predict from the early citation history of a paper whether it will prove to be great (\textit{fast-hi}) or not. Random guessing would give a 50\% success rate at this task. Figure~\ref{SuccessRateVsTrainingBinary} shows the 2-state classification success, precision and recall as a function of training time. Interestingly, there is a dip in the prediction success over the first few years, just because of the higher frequency of papers in the NOT FAST cluster and the very short training periods predicting the overwhelming majority are in the NOT FAST cluster (93.3\% in the most extreme case). The figure shows that around half of the \textit{fast-hi} papers can be identified with 16 to 17 years of training data (recall of 50\%), a comparable precision of 50\% would require 19 years of training data. 
	
	\section{Citers are losing interest in papers faster than before}
	
\begin{figure}
		\includegraphics[width=\linewidth]{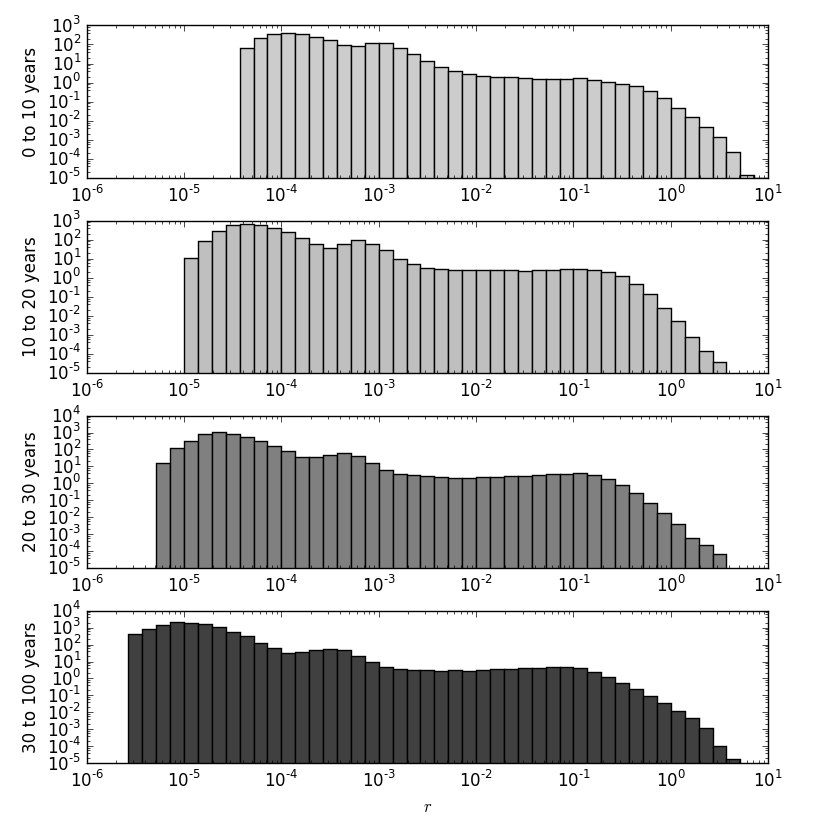}
		\caption{The distribution of the decay parameter $r$ for different age groups of papers. It is notable that the younger the papers are the more right-shifted the distribution of $r$.}
		\label{rDistributionAndAge}
	\end{figure}

	The number of papers in the scientific literature increases exponentially \cite{deSollaPrice1963}. A consequence is that interest in any given paper should naturally drop off faster in newer literature than in older literature \cite{ParoloPGHKF15}.  We observe this phenomena in our analysis. We collected papers into age groups of 0 to 10, 10 to 20 and 30 to 100 years old. We compared the distribution of $r$ for each of them; see Figure~\ref{rDistributionAndAge}. We find that the older papers (lower panels) have minimal $r$ values shifted to the left, indicating that citations diminish faster for newer papers than for older ones.
	
	\section{This model gives metrics for comparing journals}
	
	We looked at the various numbers of papers in each category in different physics subjournals. We found an enrichment of \textit{fast-hi} papers in Phys. Rev. B (PRB), Phys. Rev. Lett., and Rev. Mod. Phys. relative to the other journals. PRB has significantly higher enrichment in the number of FAST papers compared to PRA, PRC, PRD and PRE, which are not enriched (see appendix~\ref{Enrichment}). Figure~\ref{Hot2NotRatio} shows the fraction of papers in \textit{fast-hi} cluster. Journals are order from highest to lowest impact factor (left to right) respectively.
	The fraction of \textit{fast-hi} papers is significantly higher for Rev. Mod. Phys. compared to the other journals, and Phys. Rev. B and Phys. Rev. Letters are comparable. In this regard, our categorizations are consistent with PRL being high impact, RMP being a major review journal and PRB, a journal for condensed matter physics, which is the fastest growing subfield of physics in numbers of Ph.D.s awarded recently, nearly double that of its closest competitor particle physics \cite{Mulvey2011,Mulvey2014,AIPStatisticalResearchCenter2016}.
	
	\begin{figure}
		\includegraphics[width=\linewidth]{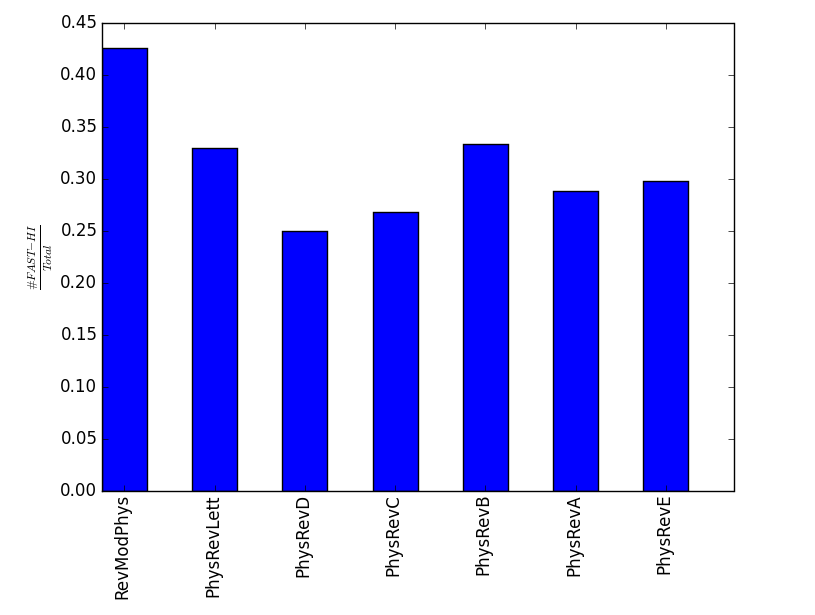}
		\caption{A bar plot comparing the fraction of papers in the \textit{fast-hi} cluster of the most important journals in Physical Review ordered by highest to lowest impact factor.}
		\label{Hot2NotRatio}
	\end{figure}
	
	\section{Conclusion}
	Using a very simple model based on the pure-birth stochastic process, we characterize the typical behavior of papers, such as a "rich-get-richer" effect and a longer term decay in citation, without too many parameters. There is the behavior of sleepers which is not captured properly by the model, such as sleepers' have unreasonably large values of parameter $a$, but this does not prevent us from utilizing the model to categorize papers based on where they are located in parameter space. There are three categories which papers cluster into high impact (\textit{fast-hi}), rapidly go obsolete (\textit{fast-flat}) and late bloomers/sleeping beauties \cite{KeRadicchi2015} (\textit{slow-late}). We saw that Rev. Mod. Phys., Phys. Rev. Lett., and Phys. Rev. B have a higher proportion of \textit{fast-hi} papers which correlates with the impact factor and yearly number of Ph.D.s awarded. The categorization of papers becomes more accurate with the more years of data available, but after 5 years of training data more than half of the papers will be correctly identified.

	\bibliography{PublicationCitation}
	
	\appendix

	\section{Methods}
	
	 We only included papers from the Physical Review corpus with more than 10 citations, to avoid over-fitting and excluded problematic papers which were cited before publication. Given data containing the times $t_1, \, t_2, \, t_3, \, \dots \, , \, t_n$ for $n$ events that occurs in a time interval from $0$ to $T$ such that $0< t_1< t_2< t_3< \dots < t_n< T$ the model parameters can be extracted by maximizing the likelihood of observing the data given the model, without needing to repeat multiple trials. This is convenient, since we will be using the birth process to describe the citations of a paper and we cannot repeat history to get multiple trials, to maximize the probability $p_{m,n}(t,\tau)$. Knowing the probability of no events in a time interval $\tau$ given $m$ events occurred in a time $t$,
	 \begin{equation}
		 p_{m,0}(t,\tau) = \exp\left( \int_{0}^{\tau} \lambda_m(t+u) \, du \right)
	 \end{equation}
	 allows us to determine the probability of any number of events occurring in the time interval $\tau$ given $m$ events occurred in a time $t$, $p_{m,n>0}(t,\tau)=1-p_{m,0}(t,\tau)$. The probability $p_{m,n>0}(t,\tau)$ is the cumulative distribution function for the waiting time distribution, since it is the chance of observing at least one event over a time interval $\tau$. Differentiating $p_{m,n>0}(t,\tau)$ with respect to $\tau$ will give the waiting time distribution or the probability that an event occurs after a time $\tau$ passes between $\tau$ and $\tau + d\tau$
	
	\begin{equation}
	f(\tau|m,t)d\tau = \lambda_m(t+\tau) \exp\left(-\int_{0}^{\tau} \lambda_m(t+u) \, du\right) \, d\tau
	\end{equation}
	so the probability density of $n$ events occurring at the times $t_1, \, t_2, \, t_3, \, \dots \, , \, t_n$ with the condition that $0< t_1< t_2< t_3< \dots < t_n< T$ is simply the products of the above formula which give the likelihood function
	
	\begin{multline}
	\mathcal{L}(a,b,\dots|t_1,t_2,t_3,\dots,t_n)\\ = f(t_1,t_2,t_3,\dots,t_n|a,b,\dots)\\
	= \exp\left( -\int_{0}^{T} \lambda_{N(t)}(t) \, dt\right) \prod_{i=1}^{n} \lambda_{i-1} (t_i)
	\label{BirthProcessLikelihood}
	\end{multline}
	Note that in the integral the rate $\lambda_{N(t)} (t)$ has an implicit time dependence on the number of events that happened up to the time $t$. Explicitly $N(t)$ can be written as
	
	\begin{equation}
	N(t) = \sum_{i=1}^n \theta(t-t_i)
	\end{equation}
	where $\theta(x)$ is the Heaviside step function which is unity for $x>0$ and zero $x<0$. The formula in equation~\eqref{BirthProcessLikelihood} is general, as we made no assumptions about the mathematical form of the rate. It is quite easy to interpret that maximum likelihood estimate (MLE) maximizes the rates at the points in time when events occur, but minimize the rate between the times of events.
	
	The MLE fitting is done in Python using the minimize function in SciPy's optimize module. The optimization was given bounds and limited to positive values except for the parameter $a$ which is bounded from below by $1 \times 10^{-8}$ to avoid divide by zero errors. Initial guesses for the parameters were determined from the curve\_fit function in SciPy's optimize module fitting the cumulative number of citations as a function of time with the expected number of citations given by equation~\eqref{MeanEventsBirthProcess} (for $m=0$ and $t=0$).

	\section{Enrichment of the clusters in the different journals}
	\label{Enrichment}
\begin{figure}
		\includegraphics[width=0.5\textwidth]{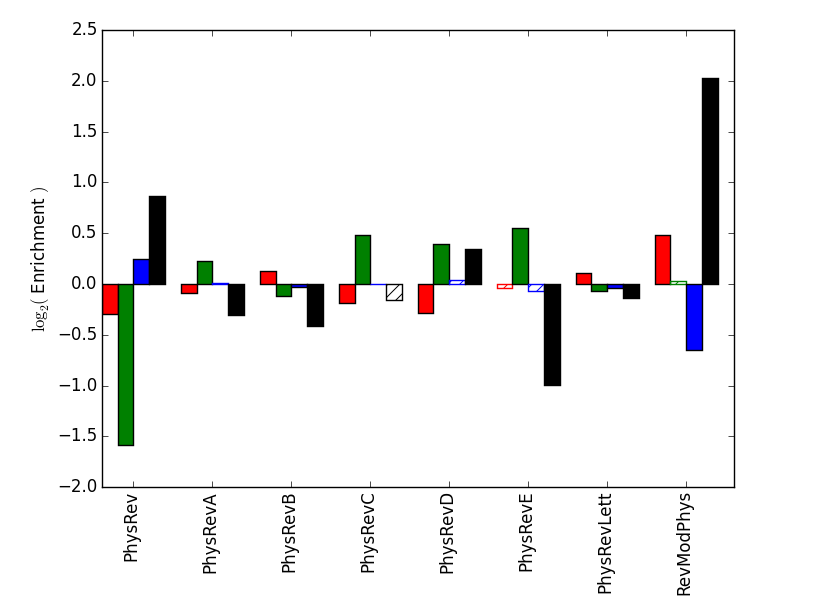}
		\caption{Logarithm of Enrichment for each journal the colors correspond to those of figure~\ref{ExpClusters}. The solid bars are statistically significant at a level of $0.01/(4 \times 12) \approx 2.1 \times 10^{-4}$ the hatched bars are not significant at that level.}
		\label{EnrichmentMajorJournals}
	\end{figure}

	Of the papers fit to our statistical model and clustered we analyzed the composition of these journals in terms of the different clusters. We took the ratio of the fraction of papers in each cluster to be the fraction of the cluster on the whole.
	We call the ratio of these two fractions the enrichment. Mathematically this is expressed as
	$E_{c, j} = \frac{f_{c, j}}{f_{c, whole}}$
	$f_{c,j}$ is the fraction of papers in cluster $c$ belonging to the journal $j$ and $f_{c,whole}$ is the fraction of papers in the cluster $c$ on the whole of the Physical Review corpus.  When the logarithm of the enrichment is positive (negative), it indicates a greater (lower) proportion of papers in that cluster for the respective journal compared to the whole.
	A value of zero means the two fractions are the same. A bar plot of the enrichments in Figure~\ref{EnrichmentMajorJournals} shows that Phys. Rev. Lett., Phys. Rev. B and Rev. Mod. Phys are the only journals to have a positive enrichment in the \textit{fast-hi} papers negative enrichment in papers from the \textit{fast-flat} cluster. 
	
	\begin{table*}
		\begin{tabular}{| l | r | r | r | r | r | r | r | r | r |}
				\hline
				Journal	& whole & \textit{fast-hi} & \textit{fast-hi} p-value & LATE & LATE p-value &	\textit{fast-flat} & \textit{fast-flat} p-value & Noise & Noise p-value \\ \hline
				PhysRev & 13682 & 3406 & 2.4e-31 & 475 & 3.2e-187 & 9142 & 1.1e-40 & 659 & 1.4e-48 \\ \hline
				PhysRevA & 15309 & 4406 & 7.1e-06 & 1866 & 2.9e-12 & 8708 & 6.7e-03 & 329 & 5.4e-06 \\ \hline
				PhysRevB & 39985 & 13326 & 1.7e-24 & 3844 & 1.6e-09 & 22020 & 4.1e-05 & 795 & 8.6e-23 \\ \hline
				PhysRevC & 7877 & 2114 & 2.5e-09 & 1149 & 1.0e-26 & 4427 & 1.2e-02 & 187 & 9.2e-03 \\ \hline
				PhysRevD & 18361 & 4595 & 1.5e-40 & 2509 & 7.3e-41 & 10640 & 3.9e-04 & 617 & 2.8e-10 \\ \hline
				PhysRevE & 6010 & 1792 & 8.8e-03 & 915 & 3.1e-26 & 3223 & 9.5e-04 & 80 & 1.1e-12 \\ \hline
				PhysRevLett & 48246 & 15922 & 6.7e-25 & 4782 & 3.0e-06 & 26376 & 7.1e-08 & 1166 & 1.0e-05 \\ \hline
				PhysRevSTAB & 99 & 24 & 6.5e-02 & 30 & 2.5e-06 & 41 & 2.5e-02 & 4 & 1.5e-01 \\ \hline
				PhysRevSTPER & 10 & 1 & 1.9e-01 & 5 & 1.2e-02 & 3 & 2.0e-01 & 1 & 2.4e-01 \\ \hline
				PhysRevSeriesI & 5 & 1 & 4.3e-01 & 0 & 6.1e-01 & 2 & 3.9e-01 & 2 & 1.6e-02 \\ \hline
				PhysRevX & 20 & 7 & 2.2e-01 & 3 & 2.3e-01 & 5 & 5.2e-02 & 5 & 4.5e-04 \\ \hline
				RevModPhys & 1478 & 630 & 1.8e-12 & 157 & 3.6e-02 & 531 & 9.4e-21 & 160 & 3.4e-46 \\ \hline
		\end{tabular}
		\caption{The number of papers in each cluster and journal compared to the whole with p-value determined by a one-sided Fisher exact test.}
		\label{JournalComposition}
	\end{table*}

\end{document}